\documentclass[letter,traditabstract,longauth]{aa}
\usepackage{graphicx}
\usepackage{txfonts}
\begin{document}
   \title{FIR colours and SEDs of nearby galaxies observed with Herschel\thanks{Herschel 
   is an ESA space observatory with science instruments provided by Principal Investigator 
   consortia. It is open for proposals for observing time from the worldwide astronomical community.}}

   \subtitle{}

  \author{A. Boselli\inst{1}
          ,
	  L. Ciesla\inst{1}
	  ,
	  V. Buat\inst{1},
	  ,
	  L. Cortese\inst{2}
	  , 
	  R. Auld\inst{2}
	  ,
	  M. Baes\inst{3}
	  ,
	  G. J. Bendo\inst{4}
	  ,
	  S. Bianchi\inst{5}
	  ,
	  J. Bock\inst{6}
	  ,
	  D.J. Bomans\inst{7}
	  ,
	  M. Bradford\inst{6}
	  ,
	  N. Castro-Rodriguez\inst{8}
	  ,
	  P. Chanial\inst{4}
	  ,
	  S. Charlot\inst{9}
	  ,
	  M. Clemens\inst{10}
	  ,
	  D. Clements\inst{4}
	  ,
	  E. Corbelli\inst{5}
	  ,
	  A. Cooray\inst{11}
	  ,
	  D. Cormier\inst{12}
	  ,
	  A. Dariush\inst{2}
	  ,
	  J. Davies\inst{2}
	  ,
	  I. De Looze\inst{3}
	  ,
	  S. di Serego Alighieri\inst{5}
	  ,
	  E. Dwek\inst{13}
	  ,
	  S. Eales\inst{2}
	  ,
	  D. Elbaz\inst{12}
	  ,
	  D. Fadda\inst{14}
	  ,
	  J. Fritz\inst{3}
	  ,
	  M. Galametz\inst{12}
	  ,
	  F. Galliano\inst{12}
	  ,
	  D.A. Garcia-Appadoo\inst{15}
	  ,
	  G. Gavazzi\inst{16}
	  ,
	  W. Gear\inst{2}
	  ,
	  C. Giovanardi\inst{5}
	  ,
	  J. Glenn\inst{17}
	  ,
	  H. Gomez\inst{2}
	  ,
	  M. Griffin\inst{2}
	  ,
	  M. Grossi\inst{18}
	  ,
	  S. Hony\inst{12}
	  ,
	  T.M. Hughes\inst{2}
	  ,
	  L. Hunt\inst{5}
	  ,
	  K. Isaak\inst{2,19}
	  ,
	  A. Jones\inst{20}
	  ,
	  L. Levenson\inst{6}
	  ,
	  N. Lu\inst{6}
	  ,
	  S.C. Madden\inst{12}
	  ,
	  B. O'Halloran\inst{4}
	  ,
	  K. Okumura\inst{12}
	  ,
	  S. Oliver\inst{21}
	  ,
	  M. Page\inst{22}
	  ,
	  P. Panuzzo\inst{12}
	  ,
	  A. Papageorgiou\inst{2}
	  ,
	  T. Parkin\inst{23}
	  ,
	  I. Perez-Fournon\inst{8}
	  ,
	  D. Pierini\inst{24}
	  ,
	  M. Pohlen\inst{2}
	  ,
	  N. Rangwala\inst{17}
	  ,
	  E. Rigby\inst{25}
	  ,
	  H. Roussel\inst{9}
	  ,
	  A. Rykala\inst{2}
	  ,
	  S. Sabatini\inst{26}
	  ,
	  N. Sacchi\inst{26}
	  ,
	  M. Sauvage\inst{12}
	  ,
	  B. Schulz\inst{27}
	  ,
	  M. Schirm\inst{23}
	  ,
	  M.W.L. Smith\inst{2}
	  ,
	  L. Spinoglio\inst{26}
	  ,
	  J. Stevens\inst{28}
	  ,
	  S. Sundar\inst{9}
	  ,
	  M. Symeonidis\inst{22}
	  ,
	  M. Trichas\inst{4}
	  ,
	  M. Vaccari\inst{29}
	  ,
	  J. Verstappen\inst{3}
	  ,
	  L. Vigroux\inst{9}
	  ,
	  C. Vlahakis\inst{30}
	  ,
	  C. Wilson\inst{23}
	  ,
	  H. Wozniak\inst{31}
	  ,
	  G. Wright\inst{27}
	  ,
	  E.M. Xilouris\inst{32}
	  ,
	  W. Zeilinger\inst{33}
	  ,
	  S. Zibetti\inst{34}
         }

\institute{	
		Laboratoire d'Astrophysique de Marseille, UMR6110 CNRS, 38 rue F. Joliot-Curie, F-13388 Marseille France
         \and 
		School of Physics and Astronomy, Cardiff University, Queens Buildings The Parade, Cardiff CF24 3AA, UK
	 \and
		Sterrenkundig Observatorium, Universiteit Gent, Krijgslaan 281 S9, B-9000 Gent, Belgium
	 \and
		Astrophysics Group, Imperial College, Blackett Laboratory, Prince Consort Road, London SW7 2AZ, UK
	 \and
		INAF-Osservatorio Astrofisico di Arcetri, Largo Enrico Fermi 5, 50125 Firenze, Italy 
	 \and
		Jet Propulsion Laboratory, Pasadena, CA 91109, United States; Department of Astronomy, California Institute of Technology, Pasadena, CA 91125, USA
	 \and
		Astronomical Institute, Ruhr-University Bochum, Universitaetsstr. 150, 44780 Bochum, Germany 
	 \and
		Instituto de Astrofísica de Canarias, C/Vía Láctea s/n, E-38200 La Laguna, Spain
	 \and
		Institut d'Astrophysique de Paris, UMR7095 CNRS, Universit\'e Pierre \& Marie Curie, 98 bis Boulevard Arago, F-75014 Paris, France
	 \and
		INAF-Osservatorio Astronomico di Padova, Vicolo dell'Osservatorio 5, 35122 Padova, Italy
	 \and
	 	Department of Physics \& Astronomy, University of California, Irvine, CA 92697, USA
	 \and
		Laboratoire AIM, CEA/DSM - CNRS - Universit\'e Paris Diderot, Irfu/Service d'Astrophysique, 91191 Gif sur Yvette, France
	 \and
		Observational  Cosmology Lab, Code 665, NASA Goddard Space Flight  Center Greenbelt, MD 20771, USA
	 \and
		NASA Herschel Science Center, California Institute of Technology, MS 100-22, Pasadena, CA 91125, USA 
	 \and
		ESO, Alonso de Cordova 3107, Vitacura, Santiago, Chile 
	 \and
	 	Universita' di Milano-Bicocca, piazza della Scienza 3, 20100, Milano, Italy 
	 \and
		Department of Astrophysical and Planetary Sciences, CASA CB-389, University of Colorado, Boulder, CO 80309, USA
	 \and
		Centro de Astronomia e Astrof\'isica da Universidade de Lisboa, Observat\'orio Astron\'omico de Lisboa, Tapada da Ajuda, 1349-018, Lisboa, Portugal
	 \and
	 	ESA Astrophysics Missions Division, ESTEC, PO Box 299, 2200
AG Noordwijk, The Netherlands
	 \and
		Institut d'Astrophysique Spatiale (IAS), Batiment 121, Universite Paris-Sud 11 and CNRS, F-91405 Orsay, France 
	 \and
		Astronomy Centre, Department of Physics and Astronomy, University of Sussex, UK
	 \and
		Mullard Space Science Laboratory, University College London, Holmbury St Mary, Dorking, Surrey RH5 6NT, UK
	 \and
	 	Dept. of Physics \& Astronomy, McMaster University, Hamilton, Ontario, L8S 4M1, Canada
	 \and
	 	Max-Planck-Institut fuer extraterrestrische Physik, Giessenbachstrasse, Postfach 1312, D-85741, Garching, Germany
	 \and
		School of Physics \& Astronomy, University of Nottingham, University Park, Nottingham NG7 2RD, UK
	 \and
		INAF-Istituto di Astrofisica Spaziale e Fisica Cosmica, via Fosso del Cavaliere 100, I-00133, Roma, Italy 
	 \and
		Infrared Processing and Analysis Center, California Institute of Technology, Mail Code 100-22, 770 South Wilson Av, Pasadena, CA 91125, USA
	 \and
		Centre for Astrophysics Research, Science and Technology Research Centre, University of Hertfordshire, College Lane, Herts AL10 9AB, UK
	 \and
		University of Padova, Department of Astronomy, Vicolo Osservatorio 3, I-35122 Padova, Italy
	 \and
		Leiden Observatory, Leiden University, P.O. Box 9513, NL-2300 RA Leiden, The Netherlands 
	 \and
		Observatoire Astronomique de Strasbourg, UMR 7550 Université de Strasbourg - CNRS, 11, rue de l'Universit\'e, F-67000 Strasbourg
	 \and
		Institute of Astronomy and Astrophysics, National Observatory of Athens, I. Metaxa and Vas. Pavlou, P. Penteli, GR-15236 Athens, Greece 
	 \and
		Institut für Astronomie, Universität Wien, Türkenschanzstr. 17, A-1180 Wien, Austria
	 \and
	 	Max-Planck-Institut fuer Astronomie, Koenigstuhl 17, D-69117 Heidelberg,  Germany 	
}

   \date{}

 
  \abstract
  {We present infrared colours (in the 25-500 $\mu$m spectral range) and UV to radio continuum spectral energy distributions of a sample of 
  51 nearby galaxies observed with SPIRE on Herschel.
  The observed sample includes all morphological classes, from quiescent ellipticals to active starbursts. 
  Active galaxies have warmer colour temperatures than normal spirals. In ellipticals hosting a radio galaxy, the far-infrared (FIR) emission is dominated by
  the synchrotron nuclear emission. The colour temperature of the cold dust is higher in quiescent E-S0a than in star-forming systems
  probably because of the different nature of their dust heating sources (evolved stellar populations, X-ray, fast electrons) and dust grain properties.
  In contrast to the colour temperature of the warm dust, the $f350/f500$ index sensitive to the cold dust decreases with star formation 
  and increases with metallicity, suggesting an overabundance of cold dust or an emissivity parameter $\beta<$2 in low metallicity, active systems.}
   {}
   {}
   {}
   {}
   {}

   \keywords{Galaxies: ISM; spiral; elliptical and lenticular; Infrared: galaxies
               }
	\authorrunning{Boselli et al.}
   \maketitle
%

\section{Introduction}

The energetic output of any extragalactic source can be determined by constructing its spectral energy distribution (SED).
The stellar component emits in the ultraviolet (UV) to near-infrared (NIR) domain, young and massive stars dominating the ultraviolet UV\footnote{The UV
emission of early-type galaxies is however due to an old stellar population (UV upturn; e.g. O'Connell 1999; Boselli et al. 2005)} 
and old stars the NIR. Dust, produced by the aggregation of metals injected into the interstellar medium (ISM) by massive stars through stellar winds and supernovae,
efficiently absorbs the stellar light, in particular that at short wavelengths, and re-emits it in the infrared domain (5$\mu$m-1mm).
At longer wavelengths, the emission of normal galaxies is generally dominated by the loss of energy of relativistic electrons 
accelerated in supernovae remnants (Lequeux 1971; Kennicutt 1983) (synchrotron emission).
Reconstructing SEDs is thus of fundamental importance for quantifying the relative contribution of the different emitting sources 
to the bolometric emission of galaxies and studying the physical relations between the various galaxy components (e.g., interstellar radiation field, 
metallicity, dust and gas content, magnetic fields). In particular, the importance of the infrared domain explored by Herschel
resides in the dust that, by means of the absorption and scattering of UV, optical and NIR photons, modifies the stellar spectra of galaxies.
SEDs are thus crucial for quantifying dust extinction and reconstructing the intrinsic distribution of the different stellar populations
within galaxies. Furthermore, fitting infrared SEDs is necessary for measuring the dust properties such as mass, temperature, fraction of PAHs,
and hardness of the interstellar radiation field (ISRF), all crucial ingredients in the study of the physical processes within the ISM
(e.g., Draine et al. 2007).\\
The interpretation of the infrared SEDs of normal galaxies has already been the subject of
several studies (Dale \& Helou 2002; Dale et al. 2001; 2005; 2007; Chary \& Elbaz 2001) even within the Virgo cluster region 
(Boselli et al. 1998; 2003). These however were generally limited in the infrared to $\lambda$ $\leq$ 170 $\mu$m, the 
spectral domain covered by ISO or Spitzer. These new Herschel data allow us to extend, for the first time for such a large sample of
normal galaxies, to the sub-mm domain ($\lambda$ $\leq$ 500 $\mu$m) where the emission is dominated by the coldest dust component.
This unexplored spectral range is crucial for determining the total mass of dust and for an accurate determination
of the total infrared luminosity. This paper presents the first, Herschel-based 
statistical study of the properties of the SED of a sample of nearby, normal galaxies
spanning a large range of morphological type and luminosity.
  
\section{The data}

Galaxies analysed in this work were observed during the Herschel (Pilbratt et al. 2010) SPIRE (Griffin et al. 2010)
science demonstration phase as part of the Herschel Reference Survey (HRS),
a guaranteed time key project designed to observe with SPIRE a volume-limited, K-band-selected, complete sample of
nearby galaxies (Boselli et al. 2010), and the Herschel Virgo Cluster Survey (HeViCS), an open time key project
focused on covering 60 sq.deg. of the Virgo cluster with PACS and SPIRE (Davies et al. 2010). To these, we added
two galaxies of the Very Nearby Galaxy Sample, M81 (Bendo et al. 2010) and M82 (Roussel et al. 2010). 
The present sample is thus composed of 51 objects with photometric 
data in the three SPIRE bands out of which 33 are Virgo members, 13 background, 3 isolated and 2 (M81 and M82) nearby galaxies.\\
The Herschel data used in this work were processed using the Level 1 procedures described in Pohlen et al. (2010),
fluxes being multiplied by a factor of 1.02, 1.05, and 0.94 at 250, 350, and 500 $\mu$m respectively, to take into account the updated
flux calibrations. Integrated flux densities were extracted
using the QPHOT task of IRAF. We assume a conservative uncertainty in the flux density of $\leq$ 30\%. This includes uncertainties on the 
absolute flux calibration (15 \%; Swinyard et al. 2010), uncertainties introduced by the map-making technique and the possible contamination of background objects
(on and off source), which might affect the flux extraction procedures. Independent observations of the galaxies NGC 4438 and NGC 4435 
(the first one being an interacting system with extended morphology, the second one a point-like early-type source)  
performed during both the HRS and the HeViCS surveys, provided consistent results to within 10 \%. This should thus be considered as an upper limit  
to the uncertainty introduced by map making and flux extraction. Absolute flux calibration uncertainties, being systematic in the three bands, 
do not affect the observed trends in the SPIRE colour-colour diagrams. 
The dataset analysed here includes SPIRE data and measurements available at other frequencies, from UV to radio centimetre. Most of these data were
taken from the GOLDMine database (Gavazzi et al. 2003). 
\\
Despite this sample not being complete in any sense, and being dominated by Virgo cluster galaxies for which perturbations 
induced by the cluster environment may lead to systematic differences in the emission properties relative to similar isolated objects even 
in the FIR spectral domain (Boselli \& Gavazzi 2006; Cortese et al. 2010), this is the first sample 
observed with Herschel that is suitable for a statistical analysis since it consists of well-known nearby 
galaxies spanning a wide range of both morphological type and luminosity.


\section{Far infrared colours}

\begin{figure}
   \centering
   \includegraphics[width=9.5cm]{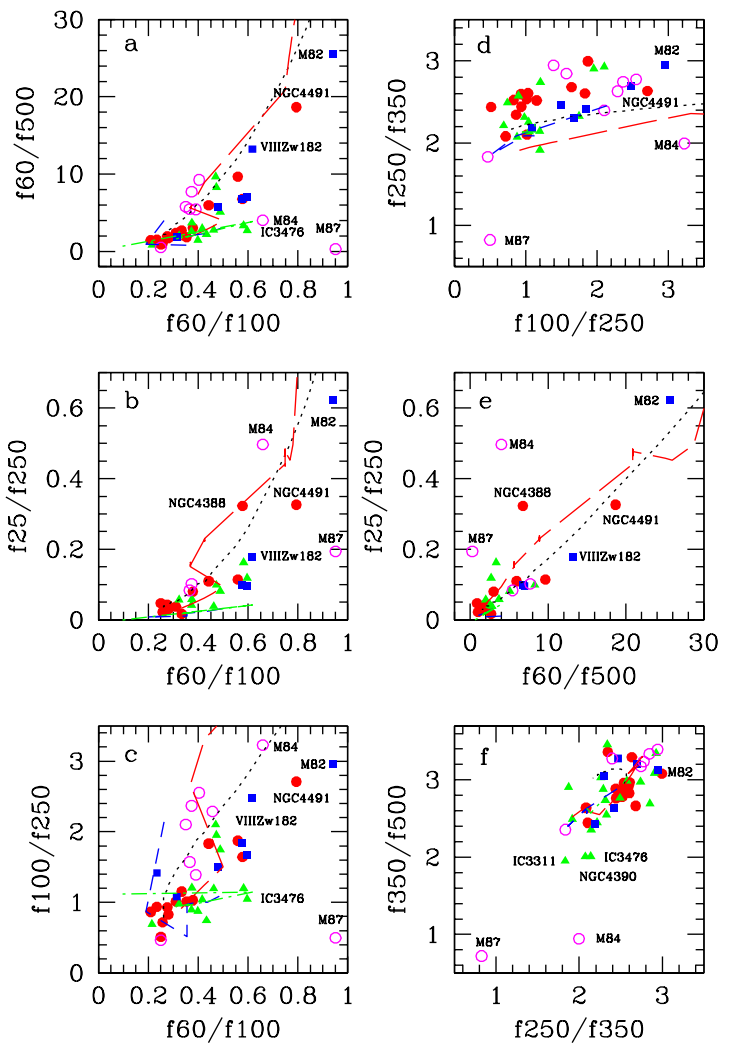}
   \caption{The infrared colours of our targets. Galaxies are coded according to their morphological type: magenta empty circles for 
   E-S0a, red filled circles for Sa-Sb, green triangles for Sbc- Scd, blue squares for Sd, Im, BCD, and Irr galaxies. The black dotted line 
   indicates the colour expected from the Dale \& Helou (2002) empirical SED, the red long-dashed line those from Chary \& Elbaz (2001), 
   the blue-short dashed, and the green dashed-dotted line the colours of the morphology- and luminosity-dependent
   templates of Boselli et al. (2003). }
   \label{colours}%
   \end{figure}

A phenomenological, model-independent technique for quantifying the spectral properties of galaxies is that of determining their colours.
To do this, we combine SPIRE and IRAS flux densities, the latter being sensitive to the warm dust component.
Figure \ref{colours} shows the IR colours of the sample galaxies.

\noindent
A first analysis of Fig. \ref{colours} indicates that in star-forming galaxies the flux density ratios $f60/f500$, $f25/f250$, or $f100/f250$ 
are strongly correlated with the generally used IRAS colour index $f60/f100$
(panels a, b and c). The dynamic range covered by $f60/f500$, however, is a factor of about 30 larger than that covered by
the $f60/f100$ flux density ratio, and is thus a much clearer tracer of the average temperature of the bulk of the dust component.
Starburst galaxies, generally defined as those objects with $f60/f100$ $>$ 0.5 (Rowan-Robinson \& Crawford 1989)
have $f60/f500$ spanning from $\sim$ 3 to $\sim$ 30. The prototype starburst galaxy in the local universe M82
has a $f60/f500$ of $\sim$ 26, significantly larger than NGC 4491, a starburst in the Virgo cluster, and VIIIZw182, a background 
merging system at $z$ = 0.07. Early spirals (Sa-Sb, red filled dots, see Table \ref{Tab}) have $f60/f500$ colours  
generally colder than Sbc-Scd (green triangles)
and Sd, Im, BCD, and Irr (blue squares). Early-types with a synchrotron-dominated IR emission (M87, M84)
are well separated in all IRAS-SPIRE or SPIRE colour diagrams with respect to the other dust-dominated E-S0a. The remaining early-types have 
colour indices indicating that the cold dust has a higher temperature than in star-forming systems.
As for the interpretation of the FIR properties of the early-type galaxies in the SINGS galaxy sample
(Draine et al. 2007), higher dust-weighted mean starlight intensities can explain the high FIR colour temperatures
of E/S0 galaxies. However, the relative importance of X-ray heating (Wolfire et al. 1995),
stochastic heating, heating from fast electrons in the hot gas, as in supernovae, and the (unknown) size-distribution of dust grains in 
these environments with low-density ISM needs further exploration. 

\begin{table}
\caption{Median colours with 1 $\sigma$ standard deviation in the colour distribution for different morphological classes}
\label{Tab}
{\scriptsize
\[
\begin{tabular}{c@{}cccccc}
\hline
\noalign{\smallskip}
Type		& f60/f100	& f250/f350	& f350/f500	& f25/f250	& f100/f250	& f60/f500 \\
\hline
E-S0a$^1$	& 0.37$\pm$0.06	& 2.60$\pm$0.38	& 3.13$\pm$0.39	& 0.09$\pm$0.01 & 1.82$\pm$0.73	& 5.69$\pm$2.96\\
Sa-Sb$^2$	& 0.34$\pm$0.12	& 2.50$\pm$0.24	& 2.86$\pm$0.22	& 0.08$\pm$0.09	& 1.10$\pm$0.42	& 3.14$\pm$2.66\\
Sbc-Scd		& 0.43$\pm$0.10	& 2.29$\pm$0.40	& 2.70$\pm$0.44	& 0.08$\pm$0.04	& 1.21$\pm$0.45	& 3.66$\pm$2.60\\
Sd-Irr		& 0.54$\pm$0.23	& 2.54$\pm$0.35	& 2.97$\pm$0.35	& 0.24$\pm$0.24	& 1.84$\pm$0.63	&10.30$\pm$9.07\\	 
\noalign{\smallskip}
\hline
\end{tabular}
\]
}
Notes: \\
1: excluding the synchrotron-dominated M84 and M87\\
2: excluding the starburst NGC 4491\\
\end{table}

\noindent
The empirical SEDs of Dale \& Helou (2002), Chary \& Elbaz (2001), and Boselli et al. (2003),
despite possible uncertainties in the absolute flux calibration (15 \%), only qualitatively cover the wide range 
of infrared colours observed in our sample even excluding the radio galaxies M87 and M84, and
underpredict the $f250/f350$ ratio for a given $f100/f250$ ratio (panel d). 
Furthermore, these models do not reproduce the coldest colour temperatures observed in the SPIRE colour 
diagram $f350/f500$ versus $f250/f350$ (panel f). It is also interesting that even the most active galaxies such 
as M82 and NGC 4491, which are expected to be dominated by warm dust heated by the dominant starburst,
host a cold dust component as traced by the 500 $\mu$m emission which is underestimated by models (see panel a). 

   \begin{figure*}
   \centering
   \includegraphics[width=13cm, angle=-90]{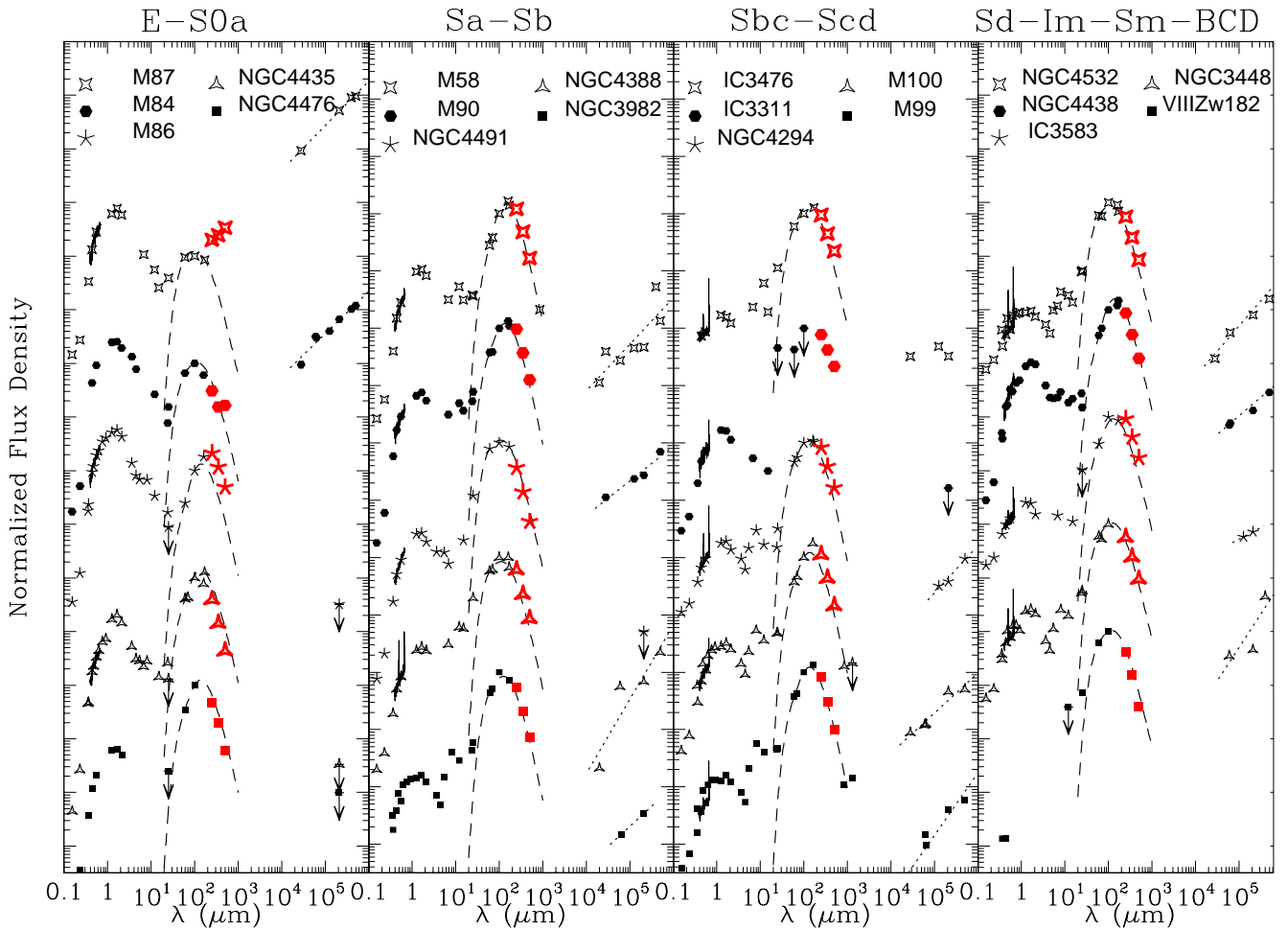}   
   \caption{The UV to radio centimetre observed SEDs of a subsample of galaxies, in 4 different panels according 
   to their morphological type. The interacting systems NGC 3448 (Arp 205) and NGC 4438 (Arp 120) are included in the Sd-BCD-Pec sample.
   SPIRE data are marked in red. The continuum line in the optical domain is the medium 
   resolution integrated spectra with emission lines in star-forming systems taken from Gavazzi et al. (2004). The dashed line shows the combination of two modified 
   black bodies ($B(\nu,T)\nu^{\beta}$ with $\beta$=2) of temperature $T$=20 and 40 K respectively normalized to the 100 and 60 $\mu$m flux densities. 
   The dotted line in the radio domain indicates the best-fit synchrotron emission.
   }
   \label{SED}%
   \end{figure*}

\section{Spectral energy distributions}

\begin{figure}
   \centering
   \includegraphics[width=8cm]{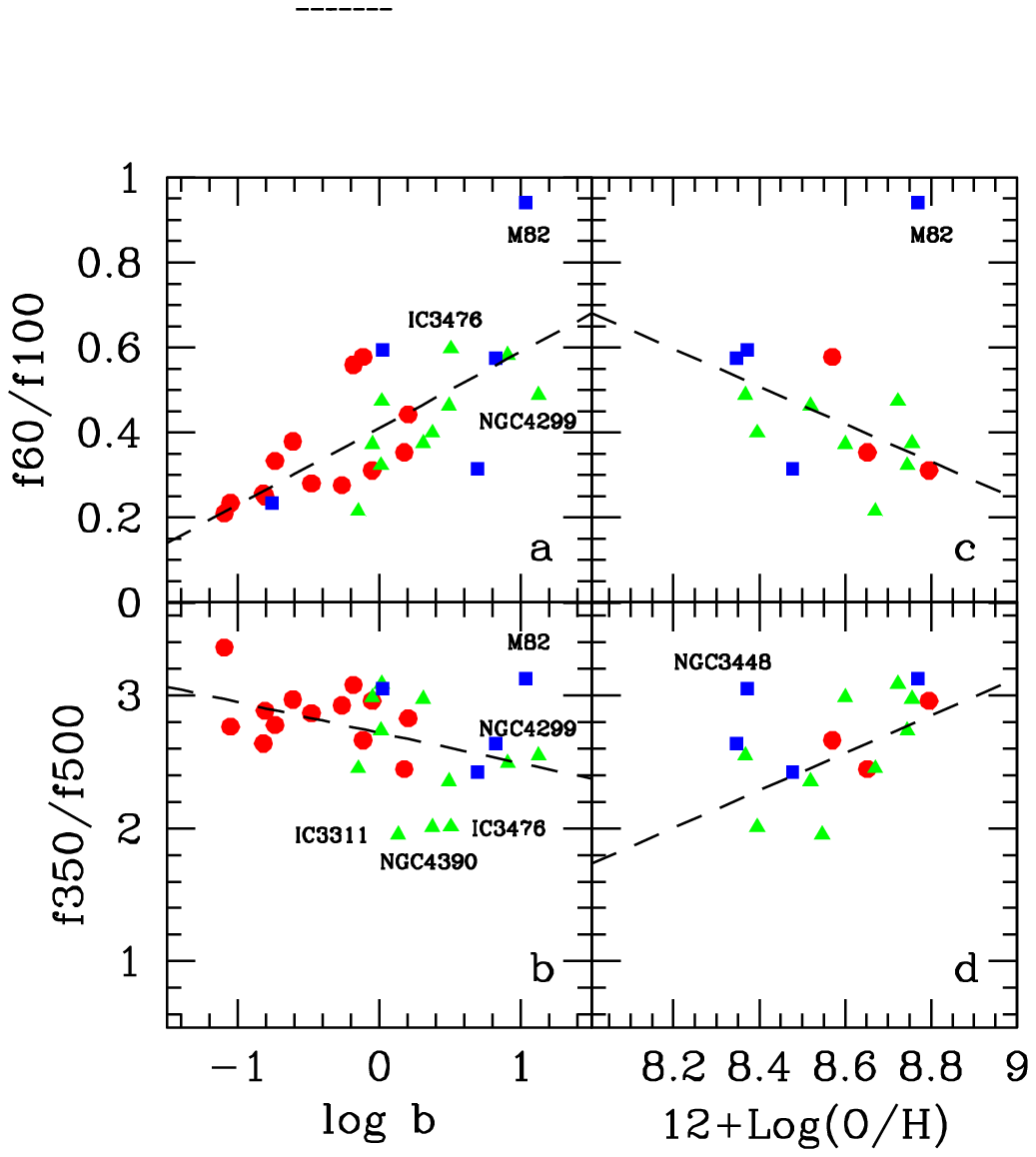}
   \caption{The relationship between the infrared colours $f60/f100$ and $f350/f500$ and the birthrate parameter $b$ (a, b)
   and the metallicity (c, d). Symbols are coded as in Fig. \ref{colours}. The dashed lines give the linear best fits to the data:
   $f60/f100$ = 0.18$\pm$0.04 log $b$ + 0.41$\pm$0.02 ($R$=0.69; $\rho$=99.99\%); $f350/f500$ = -0.23$\pm$0.10 log $b$ + 2.72$\pm$0.06 ($R$=0.39;
   $\rho$=95.16\%);
   $f60/f100$ = -0.44$\pm$0.18 [12+log(O/H)] + 4.22$\pm$1.55 ($R$=0.55; $\rho$=96.59\%); 
   $f350/f500$ = 1.41$\pm$0.55 [12+log(O/H)] - 9.56$\pm$4.70 ($R$=0.55; $\rho$=97.40\%), where
   $R$ is the correlation coefficient and $\rho$ is the Spearman's probability that the two variables are correlated. Best fits with
   metallicity were performed excluding the 2 outliers M82 and the perturbed system NGC3448 (Arp 205) since their uncertain metallicity is probably not representative of normal galaxies.
   }
   \label{SFRmetal}
   \end{figure}

Combining integrated flux densities from UV to radio centimetre we constructed the observed SED of the target galaxies.
Figure \ref{SED} shows some examples of UV to radio centimetre SEDs of galaxies according to their morphological type. 
Figure \ref{SED} shows that in the elliptical galaxies M87 and M84, the sub-mm domain is dominated by synchrotron emission.
M87 is a powerful radio galaxy (Virgo A), where synchrotron dominates down to $\sim$ 10 $\mu$m (Baes et al. 2010).
M84 is a moderately active radio galaxy with a luminosity at 20 cm of 2 $\times$ 10$^{23}$ W Hz$^{-1}$. 
In spirals, the SPIRE data closely follow a modified black body ($\beta$=2) of temperature $T$$\simeq$ 20 K (dashed line)
(e.g., Bendo et al. 2003; 2010). This however can be taken just as a first order approximation since 
quantitative data in relation to dust can be determined only after an accurate SED fitting.
%
To identify the heating sources of the emitting dust, we can use any tracer of the hardness of the
interstellar radiation field. Here we adopt the birthrate parameter $b$, defined as the ratio of the
present star formation rate (SFR) to the SFR averaged along the life $t_0$ of the galaxy\footnote{The birthrate parameter is also called
the specific star formation rate}, hence $b$ $\propto$ $SFR t_0/M_*$. Following Boselli et al. (2001, 2009), 
SFR is proportional to the extinction-corrected UV or H$\alpha$ flux, and $M_*$ to the NIR flux.
Therefore $b$ is tightly related to the hardness of the UV radiation field.
Figure \ref{SFRmetal} shows the relationship between the two colour indices $f60/f100$ and $f350/f500$ and the
birthrate parameter, this last determined for late-type galaxies only.

\noindent
Figure \ref{SFRmetal} shows that the colour index $f60/f100$, sensitive to the presence of warm dust, increases 
with $b$, indicating that galaxies with the warmest dust temperature are those at present most active in star formation
($b$ $\geq$ 1). In contrast, the temperature of the cold dust appears to be anticorrelated with
$b$, indicating an excess of the cold dust emission in the most active galaxies. These trends with $b$ may be non-universal since they might be related to
the presence of cluster galaxies, which are characterised by a reduced star formation activity ($b$ $\sim$ 0.1) because of their interaction 
with the cluster environment (e.g., Boselli \& Gavazzi 2006). An opposite behaviour of the
$f60/f100$ and $f350/f500$ colour indices is also present when plotted versus the gas metallicity
index 12+log(O/H) (determined using the prescriptions of Kewley \& Ellison (2008) based on the
Pettini \& Pagel (2004) calibration and using mainly the Gavazzi et al. 2004 spectroscopic data), i.e., while $f60/f100$ decreases with metallicity (with the exception of 
the starburst M82), $f350/f500$ seems to increase with 12+log(O/H), with a possible exception for the interacting system NGC 3448 (Arp 205). 
A similar radial trend with metallicity is also observed for both M99 and M100 (Pohlen et al. 2010).
This result is consistent with the presence of emission at $\lambda$ $>$ 850 $\mu$m
that could originate in $<$ 10 K dust (Galliano et al. 2005; Galametz et al. 2010; O'Halloran et al. 2010) or dust with $\beta$ $<$ 2 
(e.g. Bendo et al. 2006), which may be more prominent in low metallicity galaxies.
A value of $\beta$ $<$ 2 implies an enhanced contribution from carbonaceous dust because
amorphous hydrocarbons have values of $\beta$ in the range 1-1.5.
Before attempting to determine the origin of this cold dust component, this interesting result should be confirmed 
on a more robust statistical basis. 

\section{Conclusions}

Our analysis has enabled us to reach the following conclusions: a) the infrared colour index $f60/f500$ is more capable of detecting a starburst than $f60/f100$. 
b) Normal galaxies show a gradual increase in their dust temperature along the Hubble sequence, from Sa to Sc-Im-BCD with the exception of E-S0a, where the dust temperature is
higher than in star-forming systems probably because of the different nature of their dust heating sources.
c) SPIRE colours can be used to discriminate thermal from synchrotron emission in radio galaxies. d) In contrast to the warm dust, the colour temperature $f350/f500$ index 
decreases with star formation activity and increases with metallicity. This admittedly weak evidence might be indicative of an overabundance of 
cold dust or an emissivity parameter $\beta <$ 2 in low metallicity, active systems. 

\begin{acknowledgements}
SPIRE has been developed by a consortium of institutes led by
Cardiff University (UK) and including Univ. Lethbridge (Canada);
NAOC (China); CEA, LAM (France); IFSI, Univ. Padua (Italy); IAC (Spain); 
Stockholm Observatory (Sweden); Imperial College London,
RAL, UCL-MSSL, UKATC, Univ. Sussex (UK); and Caltech/JPL, IPAC,
Univ. Colorado (USA). This development has been supported by
national funding agencies: CSA (Canada); NAOC (China); CEA,
CNES, CNRS (France); ASI (Italy); MCINN (Spain); Stockholm
Observatory (Sweden); STFC (UK); and NASA (USA).

\end{acknowledgements}

\end{document}